\documentclass{aa}

\usepackage{graphicx}
\usepackage{txfonts}

\RequirePackage{natbib}
\begin{document}
\title{Forward modelling of sub-photospheric flows for time-distance helioseismology}
\author{S. Shelyag, R. Erd\'{e}lyi, M.J. Thompson}
\institute{Solar Physics and Space Plasma Research Centre [SP$^2$RC], Department
of Applied Mathematics, University of Sheffield, Hicks Building,
Hounsfield Road, S3 7RH, Sheffield, United Kingdom}
\date{01.01.01/01.01.01}

\abstract{}{}{}{}{}
\abstract
{Results of forward modelling of acoustic wave propagation in a realistic
solar sub-photosphere with two cases of steady horizontal flows are presented
and analysed by the means of local helioseismology.
}
{This paper is devoted to an analysis of the influence of steady flows on the 
propagation of sound waves through the solar interior.
}
{The simulations are based on fully compressible ideal hydrodynamical
modelling in a Cartesian grid. The initial model is characterised by solar
density and pressure stratifications taken from the standard Model S and
is adjusted in order to suppress convective instability. Acoustic waves 
are excited by a non-harmonic source located below the 
depth corresponding to the visible surface of the Sun. Numerical experiments 
with coherent horizontal flows of linear and Gaussian dependences of flow speed 
on depth are carried out. These flow fields may mimic horizontal 
motions of plasma surrounding a sunspot, differential rotation or meridional 
circulation. An inversion of the velocity profiles from the simulated travel 
time differences is carried out. The inversion is based on the ray approximation.
The results of inversion are then compared with the original velocity profiles.
}
{Results of forward modelling of acoustic wave propagation in a realistic
solar sub-photosphere with two cases of steady horizontal flows are presented. 
The influence of steady flow on the propagation of sound waves through
the solar interior is analysed. A time-distance analysis technique is applied
to compute the direct observable signatures of the background bulk motions
on travel times and phase shifts. This approach allows direct comparison
with observational data.
Further, we propose a method of obtaining the travel-time differences for the waves
propagating in sub-photospheric solar regions with horizontal flows. The method
employs directly the difference between travel-time diagrams of waves propagating
with and against the background flow.
}
{The analysis shows that the flow speed profiles obtained from inversion 
based on the ray approximation differ from the original ones. The difference 
between the original and observed profiles is caused by the fact that the wave 
packets propagate along the ray bundle, which has a finite extent, and thus 
reach deeper regions of the sub-photosphere in comparison with ray theory. 
}
 

\maketitle

\section{Introduction}

Recent observational studies in local time-distance helioseismology show 
the existence of large-scale motions of plasma in the sub-photospheric regions of 
the Sun  (see, for example, a general review of recent findings in theoretical 
and observational local and time-distance helioseismology by \citet{gizonbirch}).
These flows can be caused, for example, by interaction of the magnetic field with 
the plasma in the sunspots, by differential 
rotation, or by meridional circulation. An initial, encouraging analyses of 
sub-photospheric flows in the limit of short wavelength (ray approximation) 
has been carried out by \citet{giles}. Time-distance helioseismology in the 
ray approximation has also been used to map the flow structures beneath 
sunspots \citep{kosovichev, zhao, zhaokosovichev}. The presence
of long-lived large-scale subphotospheric convective flows has been studied
by \citet{haberetal1, haberetal2} using ring-diagram analysis. Surface gravity
waves have been used in the studies of \citet{corbard} to analyze radial gradients
of angular velocity in the solar subsurface regions down to 15~Mm. 
Progress in the analytical analysis of solar sub-photosphere with a homogeneous 
background flow has been shown by \citet{erdelyi1} and \citet{taroyan}.

The measurement accuracy of solar observations is continually changing with the
improvement of observational technology, thus requiring more precise methods of
data analysis and giving a need for artificial (synthetic) data produced by 
modelling simulations, which are able to mimic observations with a required precision. 
The approach of forward modelling of sound speed inhomogeneities in homogeneous 
solar interior models was used to explore the validity conditions of the Born and 
ray approximations by \citet{birchkosovichev}.
Also, a study of the validity of the ray and Born approximations in analyzing
the sensitivity of wave travel times to background flows, based on the forward modelling
of acoustic wave propagation in a constant background density and pressure, 
has been done by \citet{birchfelder}, \citet{gizonbirch}. 
Later, a limited approach using numerical solution of the wave equation 
with solar sound speed and density profiles has been applied to analyse seismic 
traces produced by propagation of acoustic waves, see e.g. \citet{tong}.
In a recent work of \citet{shelyagetal} the ability of full forward modelling to 
successfully reproduce the observed power spectrum of solar oscillations and seismic traces by excitation
of sound waves by single and multiple random acoustic sources has been shown. 

In this paper, an analysis is presented of the results of the forward modelling of acoustic wave propagation
in the solar sub-photosphere with two types of sub-photospheric steady horizontal 
flows. The full set of compressible hydrodynamic equations
is solved in order to provide the artificial data. The initial model is calculated
from the standard Model S \citep{jcdetal}, and is modified in order to suppress convective instability
in the originally convectively unstable solar sub-photosphere. Flows with Gaussian 
and linear velocity dependences on depth are analysed in the terms of the techniques of local 
time-distance helioseismology. The travel times and travel-time differences 
are computed from the time-dependent simulations and are compared with results calculated 
using the ray approximation.

In the next section we give a brief description of the simulation setup and outline briefly the 
numerical methods used to produce the artificial data. In Section 3, the simulations,
their direct interpretation and data measurements in terms of time-distance helioseismology 
are shown. Further, the method of obtaining travel-time differences from the 
artificial data is discussed in this section. In Section 4, we compare the 
travel-time differences obtained from forward modelling with those calculated using the ray approximation. The 
inversion technique, the flow profiles obtained by inversion of the travel-time differences, 
and their comparison with the original sub-photospheric flow velocity profiles are presented in Section 5.
The results and conclusions are discussed in Section 6.

\section{Simulation setup}
The simulation model box has been described by \citet{shelyagetal}. For convenience,
we briefly describe the setup here.
The Versatile Advection Code (VAC), originally developed by \citet{tvd1}, 
is implemented here to carry out the forward modelling. VAC solves numerically
a set of hyperbolic equations by a number of different computational methods in 
one, two or three dimensions in Cartesian, cylindrical or spherical 
geometries with uniform or non-uniform grids. The code was specifically set here to solve 
the ideal hydrodynamic equations in a two-dimensional Cartesian domain
with three-dimensional vector quantities:

\begin{equation}
\frac{\partial \rho}{\partial t} + \nabla \cdot \left(\bf{v} \rho \right)=0,
\label{hdeq1}
\end{equation}
 
\begin{equation}
\frac{\partial{\left(\rho\bf{v}\right)}}{\partial t}+\nabla\cdot\left(\bf{v}\rho\bf{v}\right)+\nabla p=\rho {\bf{g}},
\label{hdeq2}
\end{equation}
 
\begin{equation}
\frac{\partial{e}}{\partial{t}}+\nabla \cdot \left( {\bf{v}}e+{\bf{v}}p\right)=\rho {\bf{g}} \cdot {\bf{v}}+Q\left({\bf{r}},t\right),
\label{hdeq3}
\end{equation}
 
\begin{equation}
p=\left( \Gamma_1-1 \right) \left(e - \frac{\rho {\bf{v}}^2}{2}\right),
\label{hdeq4}
\end{equation}
where $\rho$ is the density, ${\bf{v}}=\left( v_x, 0, v_z \right)$ is the
two-dimensional velocity vector, $e=\epsilon+\rho {\bf{v}}^2/2$ is the total
energy density per unit volume, $\epsilon=p/\left(\Gamma_1-1\right)$, $p$ is the kinetic gas
pressure, $\Gamma_1$ is the adiabatic index, ${{\bf{g}} = \left( 0, 0, g \right)}$ is the solar
gravitational acceleration vector,
$Q$ is the temporally and spatially dependent term
describing additional energy sources, and $t$ is time.

\begin{figure}
\centering
\includegraphics[width=1.0\linewidth]{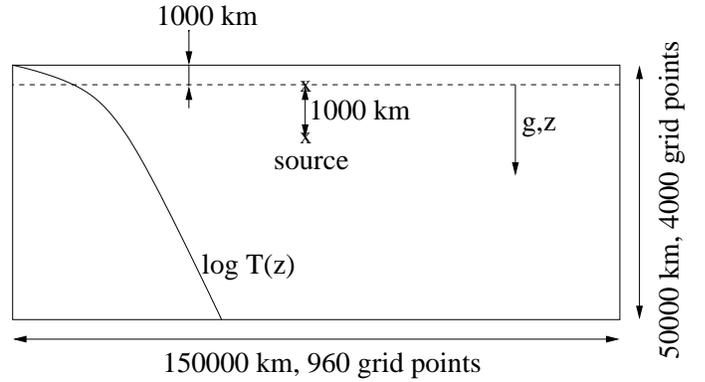}
\caption{The simulation domain. The measurement level is marked by
the dashed line. The temperature profile is also sketched on the plot.}
\label{domain}
\end{figure}

A Total Variation Diminishing (TVD) spatial and a fourth-order Runge-Kutta 
time discretisation schemes are used. The simulation domain is shown in 
Fig. 1. The box is 150 Mm wide and 50 Mm deep, and has a resolution 
of 960x4000 grid points; the upper boundary of the domain is near the 
solar temperature minimum (see Fig. \ref{domain}). The boundaries of 
the domain are open with zero gradients across the boundaries. Two boundary 
regions with $\bf{g}=0$ at the top and bottom boundaries of the box are 
introduced in order to simultaneously keep the domain in the pressure 
equilibrium and to satisfy the boundary conditions.

\begin{figure}
\centering
\includegraphics[width=1.0\linewidth]{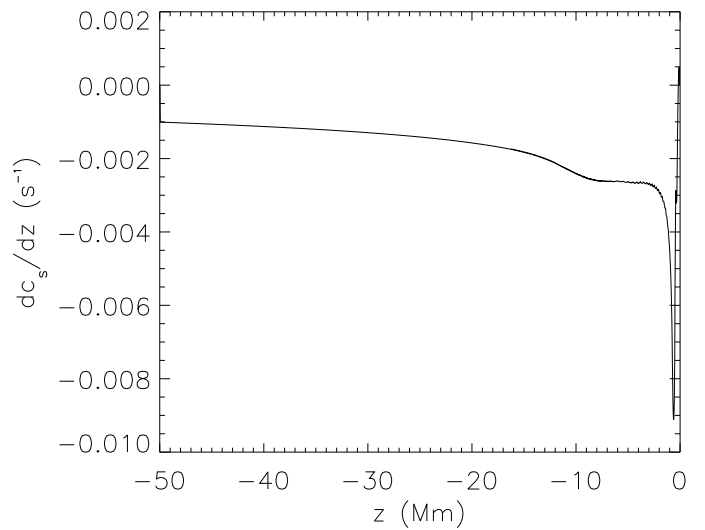}
\caption{Dependence of the gradient of the sound speed on depth in the domain.}
\label{soundspeed}
\end{figure}

The equation of state in the code is used in the form, which describes the 
properties of an ideal gas. The adiabatic index $\Gamma_1$, assumed to be 
constant over the computational domain, is equal to 5/3. The pressure 
equilibrium and modified convective stability conditions with constant solar 
gravity acceleration are applied to calculate the other quantities \citep{shelyagetal}.
The upper density value and convective instability dependence (Schwartzschild 
criterion) taken from Christensen-Dalsgaard's standard Model S, are used as 
the initial conditions for the pressure equilibrium calculation. A slight 
increase of $\Gamma_1$ in comparison to the original solar model makes most 
of the domain except the super-adiabatic strongly convective layer near the 
solar surface convectively stable. Thus, the convective instability dependence 
is modified in this layer in order to reach convective stability. This 
procedure of pressure equilibrium recalculation results in values of sound 
speed close to those in the Sun. The dependence of the sound speed gradient on depth 
in the domain is shown in Fig.~\ref{soundspeed}.

The perturbation source is located 
in the upper-middle (z=2000 km below the upper boundary) of the 
simulation box (see Fig. \ref{domain}). The measurement level for latter 
time-distance analysis is located at 1000 km below the upper boundary.

\section{Simulations}

In order to excite sound waves, a single perturbation source \citep[see][]{shelyagetal}, 
corresponding to a localised cooling event causing mass inflow and sound waves 
excitation \citep{rast}, has been introduced in the simulations. The source, 
$dQ$, is described in the energy equation of system of hydrodynamic equations as

\begin{equation}
dQ=\left[ 1+\tanh \left(\log 3 \cdot \frac{t-t_0}{\sigma_1}   \right) \right] \cdot
\exp \frac {\left( r-r_0\right)^2}{{\sigma_2}^2}.
\end{equation}

The characteristic timescale of the source is $\sigma_1=120~\mathrm{s}$, the 
spatial extent $\sigma_2= 0.5 ~\mathrm{Mm}$. The source is located in the middle of 
the horizontal layer at 2000 km depth of the simulation domain. 

\begin{figure}
\centering
\includegraphics[width=1.0\linewidth]{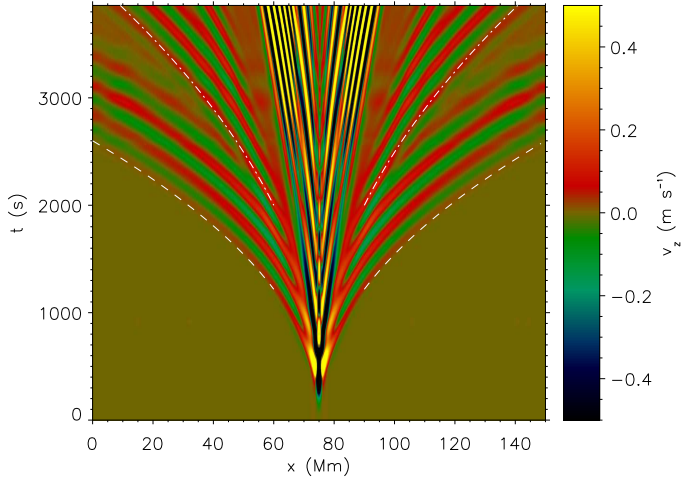}
\caption{Time-distance diagram of the acoustic response of the model 
to the source. The first and second bounces are marked approximately by dashed 
and dash-dotted lines, respectively, and are clearly visible.}
\label{vz_image}
\end{figure}

The acoustic response of the model is shown in Fig. \ref{vz_image}. The plot
shows time-distance diagram of the vertical velocity component taken at the 
measurement level of 1000 km below the upper boundary of the simulation domain. 
This level approximately corresponds to the level of the visible solar photosphere.
The first (faster) and the second (slower) bounces of the sound wave, propagating 
in the simulated solar interior, are visible in the plot. Also, the slowest wave, 
propagating in the waveguide created by the simulated temperature (and sound speed) 
minimum, is seen in the figure.

\begin{figure}
\centering
\includegraphics[width=1.0\linewidth]{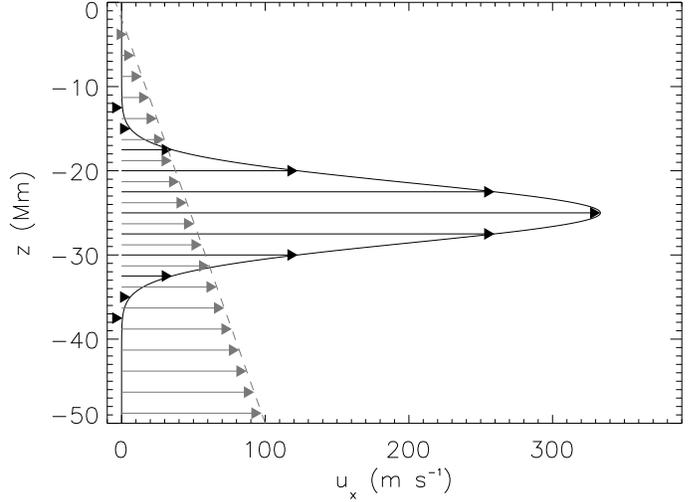}
\caption{Sub-photospheric flow structures introduced to the simulation domain.
Dashed line corresponds to the linear flow velocity profile with the maximum 
speed of $100~\mathrm{m~s^{-1}}$ at 50 Mm depth and zero at the source location, 
solid line corresponds to the Gaussian flow profile with the maximum of about 
$330~\mathrm{m~s^{-1}}$ and the FWHM of 10 Mm.}
\label{flows}
\end{figure}

We study the influence of steady flows to the acoustic response of the simulated 
solar sub-photosphere. Two types of flows are introduced (see Fig.~\ref{flows}). The first one is a 
localised flow with Gaussian horizontal velocity profile. The flow centre is 
located at the depth of 25 Mm. The width of the flow (FWHM) is 10 Mm. The maximal
flow velocity is about $300~\mathrm{m~s^{-1}}$. The flow can represent, for example, 
horizontal motion of solar plasma near a large sunspot. Existence of such mass 
flows beneath sunspots has been discovered by the means of time-distance 
helioseismology by \citet{zhao} using the ray approximation. In this work it has been shown that localised 
horizontal subsonic flows with the speeds of the order of kilometre per 
second exist at depths of about 10 Mm and have the horizontal extent of about 30 Mm 
\citep[Fig.~3 in their paper]{zhao}, the results of inversions suggest approximate cylindrical symmetry
of the flow structures under sunspots. Full magneto-hydrodynamic simulations of 
magneto-convection in convectively unstable cylindrically-symmetric polytropic 
model with presence of strong magnetic flux concentrations \citep{hurlb} also
suggest the existence of such sub-photospheric large-scale convective flows. 
These flows in the simulations appear essentially weakly magnetised in comparison to strong
magnetic flux concentrations, where convection is suppressed by magnetic field 
\citep[Fig.~2 in][]{hurlb}. 

The second case is characterised by a linear velocity dependence on the vertical coordinate 
selected in the way that the velocity at the source location is equal to zero, and 
the velocity at the bottom of the domain is $100~\mathrm{m~s^{-1}}$. This flow can 
model meridional laminar circulation. Such flows have been seen observationally by
\citet{haberetal1,haberetal2} using ring-diagram analysis of SOI-MDI observational 
data. Simulations of convection in spherical shells also show an evidence for 
meridional circulation flows with sufficiently long lifetimes \citep{elliottetal, mieschetal}.
The average dependences of the flow speed on depth for different latitudes and Carrington
longitudes, obtained by inversions of ring diagrams \citep[their Figs.~5,~6]{haberetal1},
have qualitatively similar properties as the linear flow introduced to the numerical 
experiments presented here. The flow profiles are approximately linearly dependent on depth, 
and the speeds at the depth of 15 Mm are of $25-50~\mathrm{m~s^{-1}}$.

Since the acoustic source is equidistant from the side boundaries of 
the simulation box, and without a flow the wave packets would propagate 
symmetrically around the source, one can study effects, caused by the 
Doppler shift differences between the wave packets propagating to the 
left and to the right from the source. The numerical difference between the
left and right amplitudes represents the wave phase shift.

\begin{figure}
\centering
\includegraphics[width=1.0\linewidth]{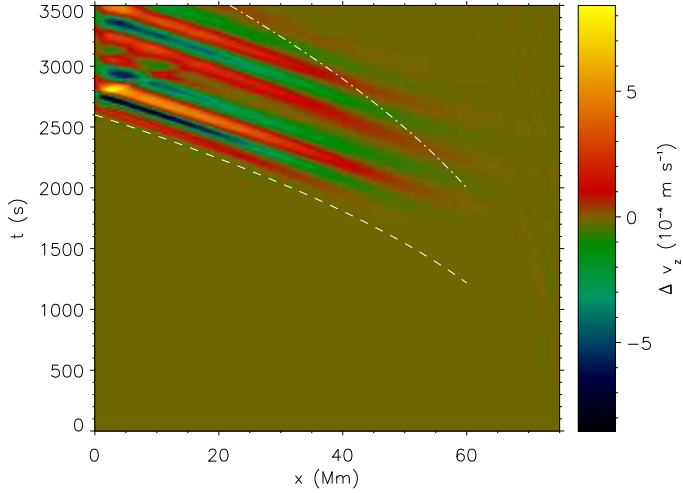}
\caption{Vertical speed difference image for the model with the Gaussian horizontal
velocity profile. The differences are computed 
between the points located at the same distance and opposite sides of 
the source. Positions of the first and second bounces in the diagram are marked by
dashed and dash-dotted lines, respectively (cf. Fig.~\ref{flows}).
The figure shows that only the first (the deepest propagating) 
bounce is influenced by the flow. Also, the flow acts on the wave 
propagating in the deeper regions with almost constant sound speed (Fig. \ref{soundspeed}), 
thus, the "difference waves" travel with a constant speed, corresponding 
to the sound speed in the region of the model, where the influence of the 
flow is most significant.}
\label{dvz_gf}
\end{figure}

The phase difference image for the case of the Gaussian flow profile is shown
in Fig.~\ref{dvz_gf}, which shows the difference between the right and left portions
of a time-distance diagram, such as Fig.~\ref{vz_image}. The introduced localised flow 
acts only on the modes which propagate
deep enough in the solar sub-photosphere; the shallow modes are much less affected.
This fact can be clearly seen from the figure. The phase differences are large
in the regions of the first bounce propagation, however, the second and higher order 
bounces and the surface wave are not influenced by the flow. Also, since the flow
is localised in the deep regions of sub-photosphere, where the sound speed is changing
slowly, the phase difference ``waves'' are propagating as the waves with
constant group speed corresponding to the local sound speed at the region where
the influence of the flow is most significant, as can be seen from the figure.

\begin{figure}
\centering
\includegraphics[width=1.0\linewidth]{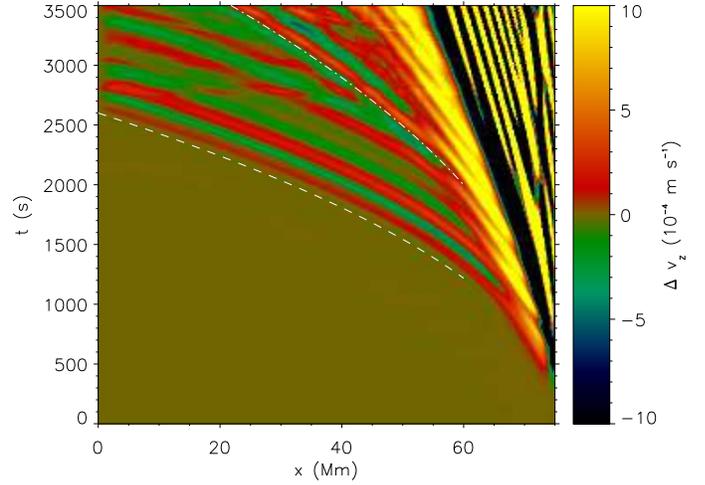}
\caption{Vertical speed difference image for the model with the linear 
horizontal velocity profile Fig. \ref{dvz_gf}. All modes in the box are influenced 
by the flow.}
\label{dvz_lf}
\end{figure}

The case for the linear horizontal speed profile is rather different. The phase difference
plot calculated for the linear profile is shown in Fig. \ref{dvz_lf}. In this case 
all of the modes are influenced by the flow.

\section{Travel-time differences}

We perform the comparison of the simulated travel times with the travel times
given by theoretical calculations using the ray approximation for the model
depth dependences of the sound speed, pressure and density used in the numerical experiments. 
The simulated travel-time differences are calculated from the phase differences 
between the wave packets propagating to the left and to the right from the source. 
Assuming the harmonic response of the model to the source, the signals to the
left and to the right of the source can be defined as 
$A_l=A_0 \sin \left(\omega \left(t-\delta \tau/2 \right) \right)$ and 
$A_r=A_0 \sin \left(\omega \left(t+\delta \tau/2 \right) \right)$, where $A_0$ is the
amplitude of the wave packet, $t$ is time, and $\delta \tau/2$ is the phase shift
of the wave packet with respect to the wave packet propagating in the model,
undisturbed by the flow. Thus, assuming that the phase differences are less 
than $2 \pi$ and measuring the difference 
$\Delta A=A_l-A_r=2 A_0 \cos \left( \omega t \right) \sin \left( \omega \cdot \delta \tau/2 \right)$
from the simulations accordingly to Figs.~\ref{dvz_gf} and \ref{dvz_lf} one
can evaluate the time difference between the acoustic waves propagating to the
left and to the right from the source:
\begin{equation}
\delta \tau=\frac{2}{\omega} \arcsin \frac{\Delta A / A}{2}.
\label{measure}
\end{equation}
In Eq.~(\ref{measure}), $A_0$ is the amplitude of the oscillation with the frequency 
$\omega$, taken along the time-distance path of the wave packet, $\Delta A$ is 
the amplitude difference between the packets propagating to the left and to 
the right from the source, taken at the same place where the amplitude $A_0$ is
taken.

In order to calculate 
the travel-time differences in the ray approximation, we follow the procedure 
described by \citet{giles}, appropriately adapted the method for the Cartesian 
geometry used in the simulations. At the solar surface, travel-time difference 
between the wave packets propagating to the left and to the right from the source 
can be written in the form:
\begin{equation}
\delta \tau = \int_{z_1}^{z_2} K \cdot u~ dz,
\label{dtaueq1}
\end{equation}
where $\delta \tau$ is the travel-time difference, $u$ is the 
horizontal flow velocity, dependent on the depth, $K$ is the integration kernel, 
which contains the information about the physics of the wave propagation 
depending on the depth. The integration limits $\left( z_1, z_2 \right)$ are
defined by the depth range where an oscillatory mode of some frequency $\omega$ 
can propagate, or, in other words, by the lower 
($z_1$) and upper ($z_2$) turning points for this mode.

In the ray approximation theory, the kernel $K$ is defined by the following 
expression:
\begin{equation}
K = 4 \frac{v_x}{v_z c^2}.
\label{kijdef}
\end{equation}

The factor 4 in Eq.~(\ref{kijdef}) indicates that in the 
case of the simulations described above, the travel-time differences for the waves 
propagating to the  left and to the right are the same, and that the time the 
acoustic mode travels from the upper turning point to the lower one is equal to 
the travel time in the opposite direction.  

In a plain-parallel model one can write the quantities under the integral as 
follows:
\begin{equation}
v_z=\frac{k_z \omega^3 c^2 }{\omega^4 - k^2_x c^2 \omega_{BV}^2},
\label{intpar1}
\end{equation}

\begin{equation}
v_x=k_x \omega c^2 \frac{\omega^2-\omega_{BV}^2}{\omega^4-k_x^2 c^2 \omega_{BV}^2},
\label{intpar2}
\end{equation}
where $\omega_{BV}$ is Brunt-V\"ais\"al\"a frequency, defined as
\begin{equation}
\omega_{BV}^2=g_z \left(\frac{1}{\Gamma_1} \frac{d \ln p}{dz} - \frac{d \ln \rho}{dz} \right),
\label{omegabv}
\end{equation}
$\omega_{AC}$ is the acoustic cutoff frequency, given by
\begin{equation}
\omega_{AC}^2=\frac{c^2}{4} \left( \frac{d \ln p}{dz} \right)^2,
\label{omegaac}
\end{equation} 
$k_z$ is vertical wavenumber, 
\begin{equation}
k_z^2=c^{-2} \left(\omega - \omega_{AC} \right)-k_x^2 \left(1-\omega_{BV}/\omega \right),
\label{kzdef}
\end{equation}
and, finally, $k_x$ is horizontal wavenumber.

The upper and lower turning points $z_1$ and $z_2$ are defined as the range where
the integrand of Eq.~(\ref{dtaueq1}) is real.

Substituting Eqs.~(\ref{intpar1})-(\ref{kzdef}) into Eq.~(\ref{kijdef}), one can 
calculate the values of the kernel, defined by Eq.~(\ref{kijdef}). However, numerical 
computations of the integral in Eq.~(\ref{dtaueq1}) with this kernel around the 
lower turning point has to be carried out carefully due to the singularity of the integrand 
there. This singularity is integrable; the procedure of integration is described 
in details by \citet{cdgt}.   

\begin{figure}
\centering
\includegraphics[width=1.0\linewidth]{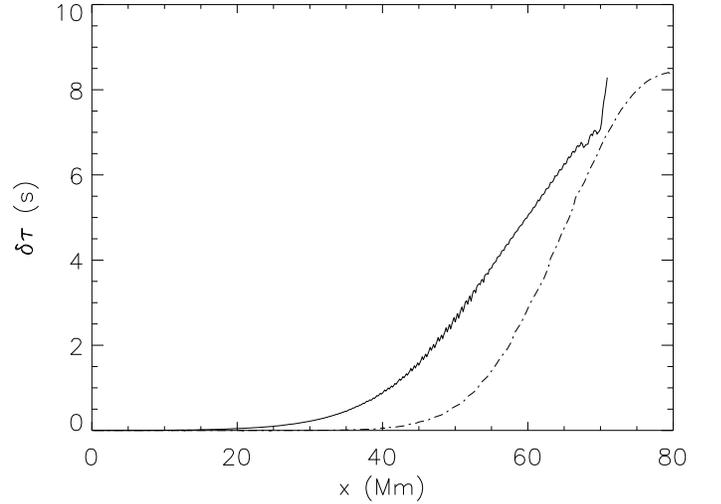}
\caption{Plot of travel-time difference for "left" and "right" propagating 
wave packets in the case of localized Gaussian flow profile. The solid curve 
corresponds to the time difference measured 
directly from the simulations, the dash-dotted curve is the time difference 
calculated using the ray approximation for density and pressure profiles 
used in the simulations. The ray approximation does not seem to give accurate results for 
small distances (and depths) in this case, however, for large distances the difference 
diminishes. The noise at distances about 70 Mm is caused by interference of
the forward-propagating wave packet with the wave reflected from the boundary
of the simulation box.}
\label{tdiff_gf}
\end{figure}

\begin{figure}
\centering
\includegraphics[width=1.0\linewidth]{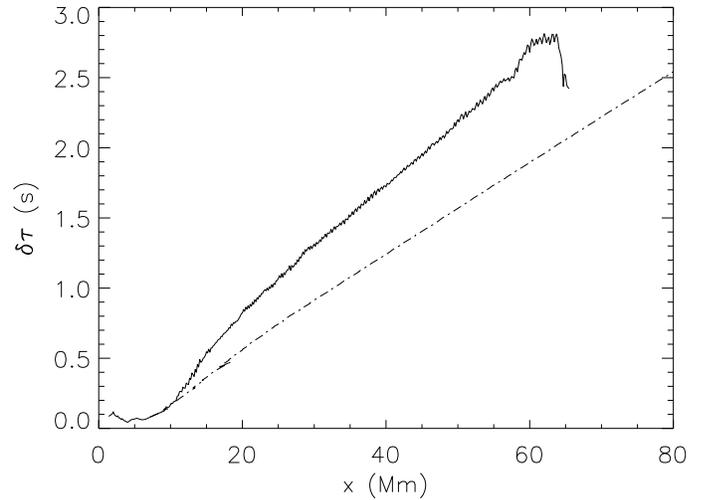}
\caption{Same as Fig. \ref{tdiff_gf}, computed for the model with linear 
velocity depth dependence. In this case the flow is not spatially localised, 
and the time difference grows with the distance from the source.}
\label{tdiff_lf}
\end{figure}

The results of the comparison are
shown in Figs. \ref{tdiff_gf} and \ref{tdiff_lf} for the Gaussian and linear horizontal 
velocity profiles, respectively. In the case of the Gaussian velocity profile 
(Fig. \ref{tdiff_gf}), the ray approximation does not give correct travel times for 
medium (30-60 Mm) distances. However, the difference between travel times derived between the 
ray approximation and the simulations diminishes on larger distances for 
the waves that propagate deep in the sub-photosphere.

The dependence of the travel-time difference on the horizontal coordinate for the 
linear velocity profile shows a rather different behaviour. All of the sound waves 
propagating in the sub-photosphere are affected by the flow, and the 
difference between the ray approximation and the forward simulations grows with the 
distance from the centre. Also, one can notice that the trend of the dependences show 
qualitatively similar behaviour.

The noise in the travel-time difference dependences at distances about
60-70 Mm from the source, is caused by the interference of the forward wave packet with the wave 
reflected from the boundary of the simulation domain, despite the reflected
wave, having much lower amplitude than the one propagating toward the boundary. 

Generally, it seems that the ray approximation does not give accurate enough travel times when compared 
with the ones derived from the simulations for the same sub-photospheric structure. This 
is caused by the fact that the wave packets, produced by superposition of $p$ 
modes, propagate along the ray bundle which follows the path given by the ray 
approximation, but has a finite extent \citep{bogdan}. Thus, for the case of 
a Gaussian flow, the wave packets actually reach deeper layers of the 
sub-photosphere and become more affected by horizontal flow, then 
predicted by the ray approximation. This difference diminishes when the 
turning point of the ray path is located close to or below the maximum of the 
horizontal flow. The wave packets, travelling in the sub-photosphere with a 
linear horizontal flow with flow speeds increasing toward the bottom of the simulation box, 
again because of their finite extent, are more influenced by the deeper and 
stronger flows in comparison with those of the ray approximation. However, as it has already
been noticed, the travel-time difference dependences are qualitatively similar:
for example, in Fig.~\ref{tdiff_lf} at e.g. the distance $x$ around 10-20 Mm,
both dependences show a similar feature. This is a change of inclination of the
curves, which is connected to the change of sound speed gradient (cf. Fig.~\ref{soundspeed})
around the depth 5-10 Mm in the sub-photosphere. This change appears 
at smaller distances for the simulations in comparison with the ray 
approximation, because the real wave packets reach the region of sub-photosphere 
with a smaller sound speed gradient earlier than the wave packets approximated 
by ray theory, due to their finite extent.

\section{Inversions of velocity profiles}

Next, we perform inversion of the flow velocity profiles in the domain using the ray
approximation. The inversions are carried out following the procedure 
described by \citet{giles}. In the case of  spatially discretized grid, Eq.~(\ref{dtaueq1}) can be 
cast in a matrix form:
\begin{equation}
K \cdot u = \delta \tau,
\label{dtaueq2}
\end{equation}
where $\tau=\tau_i$ is the vector of travel-time difference measurements, $u=u_j$ is
the vector representing the flow speed dependence on depth, $K$ is the matrix 
representation of the integration kernel Eq.~(\ref{kijdef}):
\begin{equation}
K_{ij} = 4 \int_i \frac{v_x}{v_z c^2}dz_j.
\label{kijdef1}
\end{equation}
The integral is taken along that part of the ray path which lies within the grid 
element $ij$.

The inversion problem can be set as a problem of minimisation of the quantity 
$|K u - \delta \tau|^2$ \citep[see, for example,][]{nrcpp}, which is equivalent to
\begin{equation}
K^T K u = K^T \delta \tau.
\label{dtaueq3}
\end{equation}

In general, the problem described by Eq. (\ref{dtaueq3}) can be ill-posed, also the
setup can deal with the noisy data, thus the operator $K^T K$ has to be regularised. 
In order to diminish the numerical noise coming from the simulations, we regularise
Eq.(\ref{dtaueq3}) in the following way:

\begin{equation}
\left(K^T K + \gamma \Lambda \right) u = K^T \delta \tau,
\label{dtaueq4}
\end{equation}
where $\Lambda$ is the regularisation operator, and $\gamma$ is a parameter which
controls the relative influence of regularisation on the minimisation procedure.

For the inversion problem, the system of equations Eq.~(\ref{dtaueq4}) is solved 
with respect to the flow velocity profile $u$ using the singular value decomposition 
\citep{nrcpp}. 

The forward problem of calculating $\delta \tau$, based on the given 
velocity profile and model structure, has been solved using the ray approximation. 
Then, the obtained values of $\delta \tau$ have been used in the inversion process 
in order to test the method, to select the regularisation parameter $\gamma$, and 
to compare the flow profiles obtained from the simulations and from this calculation, 
consistent with the ray approximation. The regularisation parameter $\gamma$ in 
Eq.~(\ref{dtaueq4}) is selected in such a way that it does not significantly change 
the velocity profile, obtained by inversion of the travel-time difference dependence, 
calculated using the ray approximation, in comparison with the original one. The linear 
regularisation operator of the form, describing a priori solution as a constant, is 
chosen for the inversion procedure.

\begin{figure}
\centering
\includegraphics[width=1.0\linewidth]{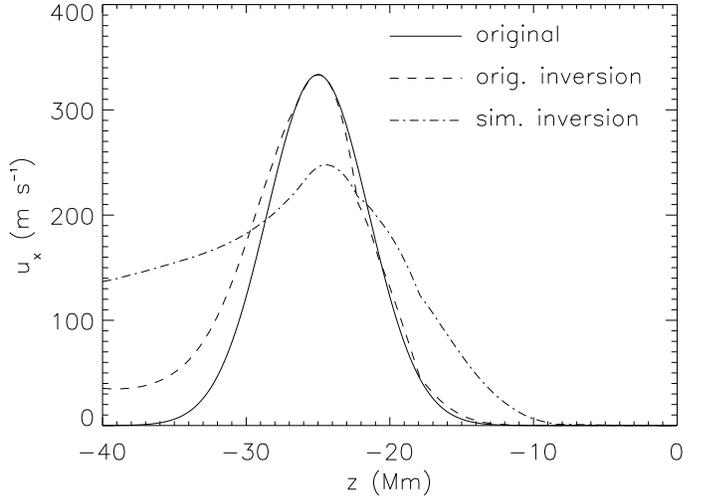}
\caption{Inversion of the velocity profile from the travel-time difference 
measurements using the ray approximation. The solid line corresponds to the original 
flow profile. The dashed line is the velocity profile calculated from 
the travel-time differences obtained by integration along the ray path given by the ray 
approximation theory for the model thermodynamic parameters. The dash-dotted line 
shows the velocity profile inferred from the travel-time difference measurements 
of the simulations using the ray approximation. The profile, inferred from the inversion, 
appears to be broader and has lower maximal amplitude than the original one.}
\label{inv_gf}
\end{figure}

\begin{figure}
\centering
\includegraphics[width=1.0\linewidth]{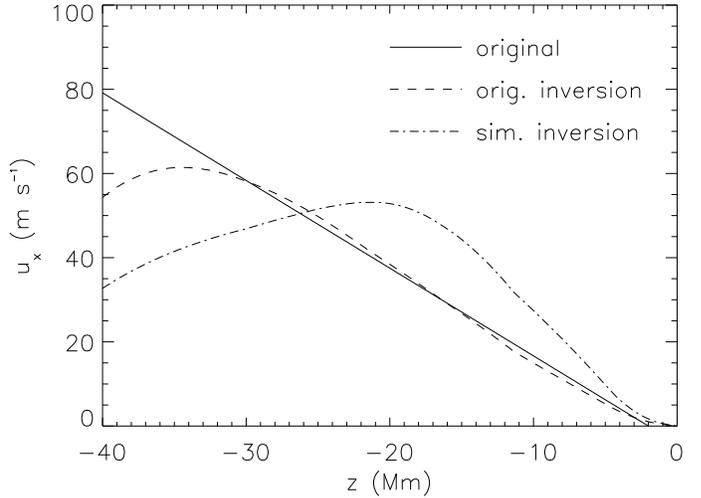}
\caption{Same as Fig. \ref{inv_gf}, calculated for the model with linear 
velocity dependence on depth. The inferred profile shows larger flow velocity 
values than the original one at depths above approximately 25~Mm, however, it
starts to decrease in the deeper sub-photospheric layers because of the
geometry of the model. }
\label{inv_lf}
\end{figure}

The results of inversion are presented in Figs.~\ref{inv_gf} and \ref{inv_lf}.
Solid curves in these plots correspond to the original flow profiles, the dashed ones
are obtained from the inversions of the travel time 
differences, which have been calculated using the ray approximation for the original 
flow profiles, and the dash-dotted curves show the flow profiles which are obtained from 
the inversions of the simulated travel-time differences (Figs.~\ref{tdiff_gf} and 
\ref{tdiff_lf}, solid curves).

In the region with approximate depth of less than 30 Mm from the solar surface, 
the original profile (Figs.~\ref{inv_gf} and \ref{inv_lf}, solid curves) and 
the profile obtained by inversion of the travel-time differences, calculated 
from the ray approximation (Figs.~\ref{inv_gf} and \ref{inv_lf}, dashed curves), 
correspond well to each other. This defines the validity range of the data 
inferred from the inversion procedure.
In the both flow cases, the results for the regions with the depth more than about 30 Mm 
do not represent the flow profiles precisely, due to the geometry of the numerical 
setup. In the model setup the wave packets that reach the deeper regions come 
out of the domain through the side boundaries and do not reach the surface.

The velocity profiles, obtained by inversions of the travel-time differences,
calculated from the simulations (Figs.~\ref{inv_gf} and \ref{inv_lf}, 
dash-dotted curves), show significantly different behaviour compared 
to the real flow profiles.
In the simulations, the wave packet, while propagating in the sub-photosphere, 
is influenced by the deeper regions of it in comparison with the ray 
approximation. Thus, in the case of horizontal flow with the Gaussian 
depth dependence, the profile, obtained from the inversion, appears to be broader, and has lower 
maximal amplitude than the original one, because the sensitivity of the
wave packet is smoothed over the range of depths. Due to a similar reason,
the velocities, obtained by the inversion, in the region between 20 and 10 Mm are larger,
than the velocities in the original profile.

In the case of the linear flow velocity profile, the velocity 
dependence, obtained by inversion, behaves in a rather different way. Above approximately 25~Mm depth 
level, the obtained profile shows velocities larger than the velocities 
of the original flow. Below 25~Mm the dependence starts to 
decrease toward the bottom of the model. Again, this is caused by the
finite width of the wave packet. A part of the wave packet leaves the domain
after it reaches the lower boundary. 

Generally, the both flow profiles are consistent with the corresponding 
travel-time difference dependencies Figs.~\ref{tdiff_gf} and \ref{tdiff_lf}.   
Larger travel-time difference corresponds to larger influence of acoustic
wave by the flow, which means the larger flow speed at the same depth level.
For the deeper regions, the decrease of the flow speed with depth
is partly due to the geometry of the simulation setup, and in the case of
Gaussian flow profile the decrease is also caused by the localised nature of the
introduced flow. 

The width of the wave packet at a particular depth can roughly be estimated 
from Fig.~\ref{inv_lf}. The width should not be less than the distance 
difference between the points on the curves of the profile obtained from 
inversion of the simulated data and the profile obtained using the ray approximation 
taken for the same value of the velocity of the flow.

\section{Discussion}

In this paper, analysis of the influence of sub-photospheric flows
on acoustic wave propagation in the solar interior is presented. 
The numerical code implemented here solves the full set of compressible hydrodynamic equations
in two dimensions. The initial model uses constant adiabatic index $\Gamma_1=5/3$ 
and solar Standard Model S 
modified in order to suppress convective instability in the solar interior,
which results in values of the sound speed close to those in the Sun.
Acoustic waves are generated by a source corresponding to photospheric
cooling events (photospheric plumes). 

Using forward modelling, the travel 
times and travel-time differences have been obtained and compared with the 
travel-time differences calculated using the ray approximation technique for
two different sub-photospheric flow profiles: localised horizontal flow with Gaussian
dependence of the flow speed on depth, and flow with a linear 
dependence of the flow speed on depth. 
The Gaussian flow may represent sub-photospheric plasma motion 
surrounding a sunspot. The linear flow may mimic large-scale flows.
The comparison of the travel-time differences shows that, in the case of a Gaussian
equilibrium motion centered about 20-30 Mm below the solar surface,
the travel-time difference between the ray approximation and forward modelling is large at small
distances from the acoustic source and diminishes at large distances. The
behaviour of acoustic waves propagating in a model with linear velocity
profile is different: there the travel-time difference grows with the distance from the source.
This difference can be explained by including some wave effects, which
are neglected in the ray approximation. Also, the discrepancy between actual travel 
times and those given by the ray approximation creates certain limitations which 
have to be taken into account in analysis of real observational data. 

The influence of the wave effects has been analysed by inversion of the
travel-time differences obtained for both cases of the 
sub-photospheric flow velocity profiles. The inversion of travel-time 
differences was carried out using the ray approximation. Then the dependences 
of the horizontal flow velocity on depth, evaluated by inversion, were compared 
with the original equilibrium bulk motion profiles. The comparison
shows that in the depth range between 0 and 20 Mm, the velocities obtained
by inversion of the modelled data 
appear to be larger than the original ones; however, in the case of Gaussian flow,
the maximal velocity is lower for the profile obtained from inversions. This is caused by the
sensitivity of the wave packet, which in reality is spread over a range of 
distances, while in the ray approximation the wave packet is presumed to be infinitely thin.

\section*{Acknowledgements}
This work was supported by the SPARG Rolling Grant of Particle Physics 
and Astronomy Research Council (PPARC, UK). RE acknowledges M.~K\'{e}ray for 
patient encouragement. RE is also grateful to NSF, Hungary (OTKA, 
Ref.No. TO43741).


\begin{thebibliography}{}


\bibitem[Birch \& Kosovichev(2000)]{birchkosovichev}
Birch, A.~C., Kosovichev, A.~G. \ 2000, Sol. Phys., 192, 193

\bibitem[Bogdan(1997)]{bogdan}
Bogdan, T.~J. 1997, ApJ, 477, 475

\bibitem[Christensen-Dalsgaard et al. (1989)] {cdgt}
Christensen-Dalsgaard, J., Gough, D.~O., Thompson, M.~J. \ 1989, MNRaS, 1989, 238, 481

\bibitem[Christensen-Dalsgaard et al.(1996)]{jcdetal} 
Christensen-Dalsgaard, J., et al.\ 1996, Science, 272, 1286 

\bibitem[Corbard \& Thompson(2002)]{corbard} Corbard, T., \& 
Thompson, M.~J.\ 2002, \solphys, 205, 211 

\bibitem[Birch \& Felder(2004)]{birchfelder} Birch, A.~C., \& 
Felder, G.\ 2004, \apj, 616, 1261 

\bibitem[Elliott et al.(2000)]{elliottetal} Elliott, J.~R., Miesch, 
M.~S., \& Toomre, J.\ 2000, \apj, 533, 546 

\bibitem[Erd{\'e}lyi \& Taroyan(2001)]{erdelyi1} Erd{\'e}lyi, 
R., \& Taroyan, Y.\ 2001, IAU Symposium, 203, 208 

\bibitem[Giles(1999)]{giles}
Giles, P.~M. Time-distance measurements of large-scale flows in the solar
convection zone. PhD thesis, Stanford University.

\bibitem[Gizon \& Birch(2005)]{gizonbirch} Gizon, L., \& Birch, 
A.~C.\ 2005, Living Reviews in Solar Physics, 2, 6 

\bibitem[Haber et al.(1998)]{haberetal1} Haber, D., Hindman, B., 
Toomre, J., Bogart, R., Schou, J., \& Hill, F.\ 1998, Structure and 
Dynamics of the Interior of the Sun and Sun-like Stars SOHO 6/GONG 98 
Workshop Abstract, June 1-4, \ 1998, Boston, Massachusetts, p.~791, 6, 791 

\bibitem[Haber et al.(2001)]{haberetal2} Haber, D.~A., Hindman, 
B.~W., Toomre, J., Bogart, R.~S., \& Hill, F.\ 2001, IAU Symposium, 203, 
211 

\bibitem[Hurlburt \& Rucklidge(2000)]{hurlb} Hurlburt, N.~E., 
\& Rucklidge, A.~M.\ 2000, \mnras, 314, 793 

\bibitem[Kosovichev et al.(2000)]{kosovichev}
Kosovichev, A.~G., Duvall, T.~L., \& Scherrer, P.~H. \ 2000, Sol. Phys., 192, 159

\bibitem[Miesch et al.(2000)]{mieschetal} Miesch, M.~S., Elliott, 
J.~R., Toomre, J., Clune, T.~L., Glatzmaier, G.~A., \& Gilman, P.~A.\ 2000, 
\apj, 532, 593 

\bibitem[Press(2002)]{nrcpp} Press, W.~H.\ 2002, Numerical 
recipes in C++ : the art of scientific computing by William 
H.~Press.~xxviii, 1,002 p.~: ill.~; 26 cm.~ Includes bibliographical 
references and index.~ISBN :  0521750334

\bibitem[{{Rast}(1999)}]{rast}
{Rast}, M.~P. 1999, ApJ, 524, 462

\bibitem[Shelyag et al.(2006)]{shelyagetal}
Shelyag S., Erd\'{e}lyi, R., Thompson, M.~J. \ 2006, ApJ, 651, 576

\bibitem[Taroyan (2004)]{taroyan}
Taroyan Y., MHD waves and resonant interactions in steady states, PhD Thesis,
The University of Sheffield, UK, 2004. 

\bibitem[Tong et al.(2003)] {tong}
Tong, C.~H., Thompson, M.~J., Warner, M.~R., Rajaguru, S.~P., \& Pain, C.~C.
\ 2003, ApJ, 582, L121

\bibitem[{{T{\'o}th} {et~al.}(1998){T{\'o}th}, {Keppens}, \& {Botchev}}]{tvd1}
{T{\'o}th}, G., {Keppens}, R., \& {Botchev}, M.~A. \ 1998, A\&A, 332, 1159

\bibitem[Zhao et al.(2001)]{zhao}
Zhao, J., Kosovichev, A.~G., \& Duvall, T.~L. \ 2001, ApJ, 557, 384

\bibitem[Zhao \& Kosovichev(2003)] {zhaokosovichev}
Zhao, J., Kosovichev, A.~G. \ 2003, ApJ, 591, 446


\end{thebibliography}
\end{document}